# E-banking and E-commerce in India and USA

Shilpan D. Vyas

School of Computer Science and Information Technology,

Singhania University, Pacheri Bari, Jhunjhunu – 333515

Rajasthan, India.

**Abstract**

Web based e-banking is becoming an important aspect of worldwide commerce. The United Nations predicts 17% of purchases by firms and individuals will be conducted online by 2006. The future of Web-based e-banking in developed areas appears bright but consumers and merchants in developing countries face in number of barriers to successful e-banking, including less reliable telecommunications infrastructure and power supplies, less access to online payment mechanisms, and relatively high costs for personal computers and Internet access. How should managers in charge of e-banking prepare for global implementation? What can they do reach consumers in developing countries? What factors influence the adoption of consumer-oriented e-banking in various countries?

This research paper will give you the idea on the local conditions in India, the Hofstede's dimension of culture in India and USA, the Diffusion of Innovation theory and hence the hypotheses for the innovation characteristics of interest.

## 1. Introduction

Much of the extent information technology (IT) literature focuses on developed countries. Less attention has been paid to IT in developing countries. As e-banking has to spread worldwide, it becomes important to understand e-banking in the context of developing countries. To better understand perception of e-banking in India, we compare them to perception of consumers in the U.S.A. by using these as a benchmark; we gain additional understanding over that which could be gained from looking at Indian consumer's perception in isolation.

## 2. Local Conditions

Local conditions such as a weak telecommunication, infrastructure, the relative expense of PCs and Internet access, weaknesses in payment infrastructure and dispute resolution systems and weaker consumer right are potential reasons for the lower penetration of e-banking in India. Liberalization in telecommunications and information technology sectors is leading to significant improvements in India's infrastructure. Also the Internet subscriber base has grown considerably (ITU, 2001). However, both personal computers and Internet access accounts are expensive in India, relative to the average income. Not having a computer at home lessens the convenience of e-banking. While the annual cost per Mbps leased line Internet connectivity in the USA is about US $ 5,000, is costs more than US $ 40,000 in India.

Worldwide, nearly two-thirds of online consumer sales place using credit cards. However, credit cards are much less common in India than in USA. While in the past there was lack of online authorization - an impediment to e-commerce and e-banking - the regulatory banking institution (the Reserve Bank of India) has released guidelines to set up "Inter-Bank Payment Gateways" to facilitate online payment, Yet for any type of distance commerce system to succeed, there needs to be a trusted dispute resolution mechanism in place. Unfortunately, India lacks such a mechanism.

Consumer rights are not a strong in India as they are in the USA. For example, "No questions asked" return policies are much less common. In e-banking transactions, consumers do not have the ability to visually inspect merchandise, which leads to increased risk. A weak dispute resolution system exacerbates this problem.

## 3. Differences in National Cultures

Difference in dimensions of national cultures may help explain differences in perception and adoption of information technologies identified five dimensions of national culture like Power distance, uncertainly avoidance, individualism vs. collectivism, masculinity vs. femininity, and long-term vs. short-term orientation. Table 1 provides definitions and index scores for the five dimension of culture for India and the USA. It also present ranking of each country as compared to the 58 countries evaluate by Hofstede, except for long-term orientation, which is based on 23 countries.

Table 1: Hofstede's dimensions of culture for India and the USA

| Dimension | Definition | | India | USA | Difference |
|---|---|---|---|---|---|
| Power Distance | Extent to which those with less power in the organizations of a country expect and accept unequal power distribution. | Score | 77 | 40 | |
| | | Rank | 10/11 | 38 | 28* |
| Individualism | Degree to which a culture reinforces individual achievement and relationships | Score | 48 | 91 | |
| | | Rank | 21 | 1 | 20* |
| Uncertainty Avoidance | Degree to which uncertain situations make members of a culture feel threatened | Score | 40 | 46 | |
| | | Rank | 45 | 43 | 2 |
| Masculinity | Degree to which distinct social gender roles characterize cultures | Score | 56 | 62 | |
| | | Rank | 20/21 | 15 | 5 |
| Long-Term Orientation** | Extent to which a culture believes in a stable society based on family, and where virtuous behavior is expected. | Score | 61 | 29 | |
| | | Rank | 7 | 17 | 10* |

* Major difference between the two countries relative to all counties evaluated by Hofstede.
** Scores for this dimension based on 23 countries versus 58 countries for the other scores.

## 4. Diffusion of Innovation Theory

Diffusion of innovation (DOI) theory is concerned with the manner in which innovations spread through social system (Rogers, 1995).

Much of the DOI research examinations on how adopter's perceptions of the characteristics of an innovation impact adoption decisions, IT innovations have been studied using this perspective. Table 2 presents the definitions of the innovation characteristics and the concept of trustworthiness.

Table 2: Definitions of the innovation characteristics and concept trustworthiness

| Characteristic | Definition | Hypotheses Indian's Vs. American |
|---|---|---|
| Relative Advantage | Degree to which an innovation is seen as being superior to its predecessor | Indians will perceive e-banking as providing less relative advantage. |
| Complexity (case of use) | Degree to which an innovation is seen as being relatively difficult to use and understand | Indians will perceive e-banking as being more complex. |
| Compatibility | Degree to which an innovation is seen as being consistent with existing beliefs values, experiences and needs | Indians will perceive e-banking being less compatible. |
| Result Demonstrability | Degree to which the outcomes of using in innovation are apparent | Indians will perceive e-commerce as providing less result demonstrability |
| Image | Degree of which the use to the innovation is seen as chancing to an individual's image or social status. | Indians will perceive e-banking as being more image chancing. |
| Trustworthiness | Degree to which consumers have confidence in the electronic marketer's reliability and integrity. | Indians will perceive Web merchants to be less trustworthy. |

# 5. Hypothesis for Innovation characteristics of Interest

Looking as differences in local conditions, the dimensions of National culture and DOI theory, we hypothesize that the subject from the two types of countries (developed and developing) differs in their perception.

Here we provide the hypothesis for this research paper work:

**Hypothesis: "Indian consumers perceive e-banking to have less relative advantages, to be less easy to use, to be less compatible, to be lower in result demonstrability, to be more image enhancing and less trustworthy than do American consumers."**

## 5.1 Indian consumers perceive e-banking to have less relative advantages than do American consumers

The convenience of shopping at home is often touted a major benefit of e-banking and e-commerce. The lower personal computer penetration rate in India may lead many Indian consumers to have to travel to a public access facility to engage in Web-based shopping. This reduces the convenience of Web-based shopping. Therefore, the relative newness of Web-based shopping may represent more of an advantage for consumers in the USA.

## 5.2 Indian consumers perceive e-banking to be less easy to use than do American consumers

First the relative unreliability of the telecommunications infrastructure in India translates into more difficulty in completing online shopping tasks. The lower use of credit cards in India may also make Web-based shopping seem more complex. If a consumer does not possess a credit card, or if a merchant cannot accept credit card payments electronically, some other method of payment must be arranged, which may lead to decreased perception of ease of use. National culture differences may also lead to difference in perceived ease of use. Communication media that do not allow for face-to-face interaction make it difficult for users in collectivist cultures to discern cues regarding the social situation that is occurring. As a result, consumers in India, which is more collectivist than the USA, may perceive online shopping to be less easy to use.

## 5.3 Indian consumers perceive e-banking to be less compatible than do American consumers

First online shopping is, in large part, a solitary activity, which may be at odds with consumers in the more collectivist culture of India. Second, American consumers are more used to engaging in "distant" shopping such as catalog shopping, which has been a fixture in the USA for many decades. Americans may also find credit cards to be more compatible with their experiences.

## 5.4 Indian consumers perceive e-banking to be lower in result demonstrability than do American consumers

In 2000, there were more than 589 Internet user per 1,000 population in USA, as opposed to 5.1 per 1,000 in India. Since the use of the Internet is much more common in the USA. The result of e-commerce should be more apparent there.

## 5.5 Indian consumers perceive e-banking to be more image enhancing than American consumers

The higher levels of Internet and personal computer use in the USA may lead to e-commerce and e-banking being more common place and thus less image enhancing.

## 5.6 Indian consumers perceive e-banking to be less trustworthy than American consumers

Trust is an important factor in determining consumer's intent to use e-commerce and e-banking. If individuals in different countries have different levels of trust, they may have different propensities to engage in e-commerce. It has been proposed that individuals in more collectivist countries are less trusting than those in more individualistic countries (Jarvenpaa &Tractinsky, 1999), which should be reflected in consumer's attitude towards Web merchants.

## 6. Conclusions

The future of banking industry depends on efforts of all concerned parties such as service providers, service facilitators, regulatory system and customers. The bank, regulatory authorities and other organizations must try their best to make banking sector as secure as possible. As regulators we need to ensure that our approaches are adequate to deal with the risks without getting in the way of the innovations and benefits that E-banking brings to firms and consumers. In USA the demand of e-banking service is more whereas it is low in India because in USA every person has e-banking services but in India only 24% people use e-banking services. So in USA banks should be introduced new technology and innovative product to attract the population. But India is progressing with a greater speed so much so that it will compare with USA in near future in e-banking business.

## References

[1] Bank for International Settlements (2003b). Risk management principles for electronic Banking.
[2] Estrella, A., Park, S., & Peristiani, S. (2000). Capital ratios as predictors of bank failures. Federal Reserve Bank of New York Economic Policy Review 6 (2), 33-52.
[3] Lin, H.-J. (2004), 'Information technology and cost and profit efficiencies in commercial banks and insurance companies: a global comparison', Unpublished Dissertation, the State University of New York at Buffalo.
[4] Teece, D. J. (1986). Profiting from technical innovation: implications for integration, collaboration, licensing and public policy. Research Policy 15, 285-306.Walter, I. (2002).
[5] Part of this article is based on an excerpt of A Brief History of Central Banking in the United States by Edward Flaherty
[6] J. Lawrence Broz; the International Origins of the Federal Reserve System Cornell University Press. 1997
[7] Carosso, Vincent P. (1973). "The Wall Street Trust from Pujo through Medina". Business History Review 47: pp. 421-437.
[8] Milton Friedman and Anna Jacobson Schwartz, A Monetary History of the United States, 1867-1960 (1963)
[9] Murray N. Rothbard. A History of Money and Banking in the United States: The Colonial Era to World War II (2002)
[10] Gupta V, (2002). "Overview of E-banking", E-banking : A Global Perspective Bankers
[11] Hawke JD (2004). "Internet Banking - Challenges for banks and regulators", Banking in the new millennium, pp 16.
[12] Malhotra and Singh, (2006). "The impact of Internet Banking on Bank's Performance: The Indian Experience", South Asian Journal of Management,Vol.13 ,No.4, pp25-54
[13] Mishra A.K. (2005). "Internet Banking in India, Part – I", www.banknetindia.com
[14] Rao N K, (2005). "Indian Banking system: The Forward March", Indian Banking 2005, The ICFAI journal of Professional Banker, February pp 7-10.
[15] www.onlinesbi.com, accessed on 25th Feb 2006
[16] http://www.internetworldstats.com
[17] An introduction to computers and their application in banking – by T.N. Srivastava, McMillan Publications The Economist 1999
[18] Banking - by Vatsala Kamat, as mentioned in article at www.outlookmoney.com
[19] K.V. Kamath, Chairman, ICICI Bank, as quoted in India Today, 27 February 2006, page 61
[20] Essentials for success with internet banking – Todd Hutto, Western Banking, Feb – Mar 2002, pg-34
[21] E-banking in India – major development and issues, by S. S. Debashish and B.P. Mishra- Pranjana, vol. 6, no.1, Jan – July 2003, pg 19
[22] E-Banking In India – Major Development and Issues", by Sathya Swaroop Debashish and Bishnu Priya Mishra – Pranjana, vol. 6, no.1. Jan – July 2003., page 18
[23] www.banknetindia.com
[24] Internet banking in India by Dr. A.K. Mishra, IIM Lucknow
[25] www.rbi.org.in

**Shilpan Vyas** is pursuing his Doctorate degree at Singhania University, Rajasthan, India. He achieved his MCA degree is 2009 at SMU, Sikkim. He is currently Lecturer at AMCOST, Faculty of Comp. Sci., Anand. His research interests are in E-banking & E-commerce, Software Testing, Data Comm. & Comp. Networks and Software Engineering.